\documentclass[prd,aps,showpacs,epsf,floats,onecolumn]{revtex4}%
\usepackage{amssymb}
\usepackage{amsfonts}
\usepackage{amsmath}
\usepackage{graphicx}%
\providecommand{\U}[1]{\protect\rule{.1in}{.1in}}
\begin{document}
\title{\textbf{Information Geometry Aspects of Minimum Entropy Production Paths from
Quantum Mechanical Evolutions}}
\author{\textbf{Carlo Cafaro}$^{1}$ and \textbf{Paul M.\ Alsing}$^{2}$}
\affiliation{$^{1}$SUNY Polytechnic Institute, 12203 Albany, New York, USA}
\affiliation{$^{2}$Air Force Research Laboratory, Information Directorate, 13441 Rome, New
York, USA}

\begin{abstract}
We present an information geometric analysis of entropic speeds and entropy
production rates in geodesic evolution on manifolds of parametrized quantum
states. These pure states emerge as outputs of suitable $\mathrm{su}\left(
2\text{; }%
\mathbb{C}
\right)  $ time-dependent Hamiltonian operators used to describe distinct
types of analog quantum search schemes. The Riemannian metrization on the
manifold is specified by the Fisher information evaluated along the
parametrized squared probability amplitudes obtained from analysis of the
temporal quantum mechanical evolution of a spin-$1/2$ particle in an external
time-dependent magnetic field that specifies the $\mathrm{su}\left(  2\text{;
}%
\mathbb{C}
\right)  $ Hamiltonian model. We employ a minimum action method to transfer a
quantum system from an initial state to a final state on the manifold in a
finite temporal interval. Furthermore, we demonstrate that the minimizing
(optimum) path is the shortest (geodesic) path between the two states, and, in
particular, minimizes also the total entropy production that occurs during the
transfer. Finally, by evaluating the entropic speed and the total entropy
production along the optimum transfer paths in a number of physical scenarios
of interest in analog quantum search problems, we show in a clear quantitative
manner that to a faster transfer there corresponds necessarily a higher
entropy production rate. Thus, we conclude that lower entropic efficiency
values appear to accompany higher entropic speed values in quantum transfer processes.

\end{abstract}

\maketitle

\section{Introduction}

Riemannian geometry has been employed in a variety of different approaches in
quantum searching \cite{grover97,nielsenbook}. In \cite{alvarez00}, it was
shown that Grover's algorithm is specified by a unitary and adiabatic process
that preserves the Fisher information function. In \cite{wadati01}, the role
of entanglement in quantum search was investigated in terms of the
Fubini-Study metric. In \cite{cafaro2012A, cafaro2012B}, quantifying the
notion of quantum distinguishability between parametric density operators by
means of the Wigner-Yanase quantum information metric, it was shown that the
quantum search problem can be recast in an information geometric framework
wherein Grover's dynamics is characterized by a geodesic on the manifold of
parametric density operators of pure quantum states constructed from the
continuous approximation of the parametric quantum output state in Grover's
algorithm. Finally, in Ref. \cite{cafaro2017}, methods of information geometry
were used to confirm the superfluity of the Walsh-Hadamard operation and, most
importantly, to recover the quadratic speedup relation.

Thermodynamic perspectives on quantum computation \cite{bennett82} and
information \cite{parrondo15}, including quantum error correction
\cite{carlo13,carlo14b}, can be quite insightful. In the framework of
classical and quantum algorithms viewed in terms of simple quantum circuit
models, the performance of search schemes is usually quantified by means of
the query complexity of the algorithm (that is, the number of oracle queries
made by the algorithm). However, more realistic models of computation should
be considered to properly analyze quantum speedups. In particular, a
physically realistic analysis should also take into account the thermodynamic
resource costs of running these algorithms on an actual computer. Some
preliminary findings are beginning to appear in the literature
\cite{beals13,perlner17}. For instance, using Bennett's Brownian model of low
power reversible computation \cite{bennett82}, Perlner and Liu argued in Ref.
\cite{perlner17} that classical exhaustive search can be quite competitive
with Grover's quantum search algorithm when the comparison between the two
searching schemes is made in terms of actual thermodynamic resource costs,
including energy consumption, memory size, and time. The comparative analysis
presented in Ref. \cite{perlner17} is mathematical in flavor and, most of all,
focuses solely on comparing classical search schemes with Grover's quantum
search algorithm. A comparison among distinct quantum searching schemes is
absent in Ref. \cite{perlner17}.

In Ref. \cite{cafaropre18}, we presented an information geometric
characterization of the oscillatory or monotonic behavior of statistically
parametrized squared probability amplitudes emerging from special functional
forms of the Fisher information function selected \emph{ad hoc}: constant,
exponential decay, and power-law decay. Furthermore, for each case, we
computed both the speed and the thermodynamic divergence of the corresponding
physical processes by exploiting a convenient Riemannian geometrization of
useful thermodynamical concepts. Finally, we briefly discussed the possibility
of employing the proposed information geometric perspective to help
characterize a convenient trade-off between speed and thermodynamic efficiency
in quantum search algorithms. A limitation of the work in Ref.
\cite{cafaropre18} is that the Fisher information functions were selected in
an \emph{ad hoc} fashion without specifying their emergence from a precise
physical setting. Therefore, despite the mathematical generality, the observed
behaviors of the parametrized squared probability amplitudes that emerged from
our information geometric analysis had no obvious physical interpretation.
Inspired by the work of Byrnes and collaborators \cite{byrnes18}, we presented
in Ref. \cite{alsing19} a detailed analysis concerning the physical connection
between quantum search Hamiltonians and exactly solvable time-dependent,
two-level quantum systems. More specifically, we analytically calculated the
transition probabilities from a source state to a target state in a number of
physical scenarios characterized by a spin-$1/2$ particle immersed in an
external time-dependent magnetic field. In particular, we investigated both
the periodic oscillatory as well as the monotonic temporal behaviors of such
transition probabilities and, additionally, explored their analogy with
characteristic features of Grover-like and fixed-point quantum search
algorithms, respectively. Finally, we discussed from a physics perspective the
connection between the schedule of a search algorithm, in both adiabatic and
nonadiabatic quantum mechanical evolutions, and the control magnetic fields in
a time-dependent driving Hamiltonian.

In this paper, motivated by the lack of any comparative thermodynamical
analysis of quantum search algorithms and building on our previous works
presented in Refs. \cite{cafaropre18,alsing19,alsing19b}, we borrow the idea
of Riemannian geometrization of the concepts of efficiency and speed within
both quantum and thermodynamical settings in order to provide a theoretical
perspective on the trade-off between speed and efficiency in terms of minimum
entropy production paths emerging from quantum mechanical evolutions.
Specifically, we present an information geometric analysis of entropic speeds
and entropy production rates in geodesic evolution on statistical manifolds of
parametrized quantum states arising as outputs of \textrm{su}$\left(  2\text{;
}%
\mathbb{C}
\right)  $ Hamiltonian models mimicking different types of continuous-time
quantum search evolutions.

The layout of the remainder of this paper is as follows. In Sec. II, we
present some preliminary information geometric concepts. Specifically, we
focus our attention on the notions of Fisher information, thermodynamic
length, and optimum paths. In Sec. III, we explain how the pure states that we
consider emerge as outputs of suitable $\mathrm{su}\left(  2\text{; }%
\mathbb{C}
\right)  $ time-dependent Hamiltonian evolutions used to describe different
types of analog quantum search schemes. In Sec. IV, we introduce the
Riemannian metrization on the parameter manifold specified by the Fisher
information evaluated along the parametrized squared probability amplitudes
obtained from analysis of the temporal evolution of a spin-$1/2$ particle in
the external time-dependent magnetic field characterizing the $\mathrm{su}%
\left(  2\text{; }%
\mathbb{C}
\right)  $ Hamiltonian model. In particular, we use a minimum action method to
transfer a quantum system from an initial state to a final state on the
manifold in finite time. Moreover, we demonstrate that the minimizing
(optimum) path is the shortest (geodesic) path between the two states, and, in
particular, minimizes also the total entropy production that occurs during the
transfer. \ Finally, by calculating the entropic speed and the total entropy
production along the optimum transfer paths in a number of physical scenarios
of interest in analog quantum search problems, we demonstrate in a
transparent, quantitative manner that to a faster transfer there corresponds
necessarily a higher entropy production rate.\textbf{ }Finally, our concluding
remarks appear in Sec. V.

\section{Information geometric preliminaries}

In Sec. II, we introduce some preliminary information geometric notions. In
particular, we discuss the concepts of Fisher information, thermodynamic
length, and optimum paths.

\subsection{Fisher information}

From an information-theoretic perspective, the concept of Fisher information
can be applied to either a multi-parameter case or a single parameter case. In
the former case, the Fisher information matrix $\mathcal{F}\left(
\theta\right)  $ with elements $\mathcal{F}_{\alpha\beta}\left(
\theta\right)  $ is defined as \cite{cover06,frieden98},%
\begin{equation}
\mathcal{F}_{\alpha\beta}\left(  \theta\right)  \overset{\text{def}}{=}%
\sum_{x\in\mathcal{X}}p_{x}\left(  \theta\right)  \frac{\partial\log\left[
p_{x}\left(  \theta\right)  \right]  }{\partial\theta^{\alpha}}\frac
{\partial\log\left[  p_{x}\left(  \theta\right)  \right]  }{\partial
\theta^{\beta}}\text{,} \label{JIJ2}%
\end{equation}
where \textquotedblleft$\log$\textquotedblright\ denotes the natural
logarithmic function. In Eq. (\ref{JIJ2}), we assume $X$ denotes a discrete
random variable with alphabet $\mathcal{X}$ and probability mass function
$p_{X}\left(  x\text{; }\theta\right)  =p_{x}\left(  \theta\right)  $.
Furthermore, $\theta\overset{\text{def}}{=}\left(  \theta^{1}\text{,...,
}\theta^{M}\right)  $ with $M$ being the dimensionality of the parameter space
$\Theta\overset{\text{def}}{=}\left\{  \theta\right\}  $. The quantity
$\mathcal{F}_{\alpha\beta}\left(  \theta\right)  $ in\ Eq. (\ref{JIJ2})\ is a
measure of the minimum error in estimating a parameter $\theta$ of a
distribution $p_{x}\left(  \theta\right)  $.\ Specifically, the Fisher
information $\mathcal{F}\left(  \theta\right)  $ obtained from $\mathcal{F}%
_{\alpha\beta}\left(  \theta\right)  $ in Eq. (\ref{JIJ2}) by assuming a
one-dimensional parameter space is formally defined as the variance of the
score $\mathrm{V}\overset{\text{def}}{=}\partial_{\theta}\left\{  \log\left[
p_{x}\left(  \theta\right)  \right]  \right\}  $ with $\partial_{\theta
}\overset{\text{def}}{=}\partial/\partial_{\theta}$, $\mathcal{F}\left(
\theta\right)  \overset{\text{def}}{=}\mathrm{var}\left(  \mathrm{V}\right)
=E_{\theta}\left[  \left\{  \partial_{\theta}\log\left[  p_{x}\left(
\theta\right)  \right]  \right\}  ^{2}\right]  $. The quantity $E_{\theta
}\left[  \cdot\right]  $ denotes the expected value of the random variable
$\mathrm{V}$ squared with respect to the probability mass function
$p_{x}\left(  \theta\right)  $. We remark that the mean value of the score is
zero. For the sake of convenience, we also point out that in the
multi-parameter case, $\mathcal{F}_{\alpha\beta}\left(  \theta\right)  $ in
Eq. (\ref{JIJ2}) can be recast as%
\begin{equation}
\mathcal{F}_{\alpha\beta}\left(  \theta\right)  \overset{\text{def}}%
{=}E_{\theta}\left[  \left(  \mathrm{V}_{\alpha}-\left\langle \mathrm{V}%
_{\alpha}\right\rangle \right)  \left(  \mathrm{V}_{\beta}-\left\langle
\mathrm{V}_{\beta}\right\rangle \right)  \right]  \text{.} \label{fifa}%
\end{equation}
In Eq. (\ref{fifa}), $\left\langle \mathrm{V}_{\alpha}\right\rangle
\overset{\text{def}}{=}E_{\theta}\left[  \mathrm{V}_{\alpha}\right]  $ and
$\mathrm{V}_{\alpha}\overset{\text{def}}{=}\partial_{\alpha}\left\{
\log\left[  p_{x}\left(  \theta\right)  \right]  \right\}  $ with
$\partial_{\alpha}\overset{\text{def}}{=}\partial/\partial\theta^{\alpha}$.
The relevance of the Fisher information is encoded in the Cramer-Rao
inequality which states that the mean-squared error of any unbiased estimator
$T\left(  X\right)  $ of the parameter $\theta$ is lower bounded by the
reciprocal of the Fisher information \cite{cover06}, $\mathrm{var}\left(
T\right)  \geq1/\mathcal{F}\left(  \theta\right)  $. Roughly speaking,
$\mathcal{F}\left(  \theta\right)  $ is a measure of the amount of information
about $\theta$ that is present in the data and gives a lower bound on the
error in estimating $\theta$ from the data.

In addition to being central to the fields of information theory
\cite{cover06} and information geometry \cite{amari00}, the concept of Fisher
information plays a key role in the geometric descriptions of both quantum
mechanics \cite{caves94,brau996} and statistical mechanics \cite{crooks12}.
For an overview of the physical meaning of the Fisher information in
information theory, quantum mechanics, and thermodynamics, we refer to Table
I. \begin{table}[t]
\centering
\begin{tabular}
[c]{c|c|c|c|c}\hline\hline
Theoretical Framework & Fluctuating Observable Quantity & Parameter of
Interest & Fisher Information & Length\\\hline
information theory & score function & elapsed time & variance of the score &
entropic\\
thermodynamics & energy & temperature & size of energy fluctuations &
thermodynamic\\
quantum theory & Hermitian operator & magnetic field intensity & dispersion of
the operator & statistical\\\hline
\end{tabular}
\caption{Schematic description of typical fluctuating observable quantities
and parameters of interest together with the physical interpretation of the
concepts of Fisher information and length through parameter space in
information theory, thermodynamics, and quantum theory.}%
\end{table}

\subsection{Thermodynamic length}

The Riemannian metric tensor introduced in\ Eq. (\ref{JIJ2}) allows one to
define the notions of length of a path and distance between two states in the
particular state space being considered. The state space may have quantum
origin or thermodynamical origin. More generally, one deals with an
information-theoretic state space where states are parametrized by parameters
more general than those used in thermodynamics. For these general scenarios,
it is customary to define the metric tensor in terms of the
information-theoretic concept of entropy where the Fisher metric tensor is
employed to define a notion of distance between points in the space of parameters.

In Ref. \cite{weinhold75}, using the second derivatives of the internal energy
with respect to extensive variables such as volume, Weinhold proposed a
Riemannian metric in the space of thermodynamic equilibrium states. In Ref.
\cite{ruppeiner79}, Ruppeiner presented a Riemannian geometric model of
thermodynamics with a Riemann structure specified in terms of a metric tensor
defined by means of second derivatives of the entropy as a function of
extensive variables such as volume and mole number. In Ref. \cite{salamon84},
Salamon and collaborators showed that Weinhold's energy version and
Ruppeiner's entropy version of the thermodynamic metric tensor are conformally
equivalent. Using the energy version of the thermodynamic metric tensor,
Salamon and Berry defined in Ref. \cite{salamon83} the length of a path
$\gamma_{\theta}$\textbf{ }with the parameter\textbf{ }$\theta$ parametrized
by an affine parameter $\xi$ with $0\leq\xi\leq\tau$ in the space of thermal
states as,%
\begin{equation}
\mathcal{L}\left(  \tau\right)  \overset{\text{def}}{=}\int_{0}^{\tau}%
\sqrt{\frac{d\theta^{\alpha}}{d\xi}g_{\alpha\beta}\left(  \theta\right)
\frac{d\theta^{\beta}}{d\xi}}d\xi\text{,} \label{length}%
\end{equation}
where $g_{\alpha\beta}\left(  \theta\right)  $ denotes the\textbf{
}thermodynamic metric tensor. The quantity $\mathcal{L}\left(  \tau\right)  $
in Eq. (\ref{length}) is known as the thermodynamic length of the path
$\gamma_{\theta}$ and has dimensions of (energy)$^{1/2}$. Clearly, it is also
possible to define $\mathcal{L}\left(  \tau\right)  $ in terms of the entropy
version of the thermodynamic metric tensor. In such a case, the corresponding
\textquotedblleft entropy\textquotedblright\ and \textquotedblleft
energy\textquotedblright\ lengths emerging from Eq. (\ref{length}) will simply
differ by a factor of the square root of some mean temperature during the
thermodynamic process being considered \cite{salamon84,salamon83}. More
generally, it is also possible to show that under suitable working conditions,
the entropy defined by a probability distribution leads to a length in the
space of probability distributions that equals the length computed using the
thermodynamic entropy in the space of extensive variables \cite{salamon85}. To
understand the physical interpretation of the thermodynamic length, it is
convenient to introduce the so-called thermodynamic divergence $\mathcal{I}%
\left(  \tau\right)  $ of a path $\gamma_{\theta}$ with the variable\textbf{
}$\theta$\textbf{ }expressed in terms of an affine parameter $\xi$ with
$0\leq\xi\leq\tau$ \cite{crooks07},%
\begin{equation}
\mathcal{I}\left(  \tau\right)  \overset{\text{def}}{=}\tau\int_{0}^{\tau
}\frac{d\theta^{\alpha}}{d\xi}g_{\alpha\beta}\left(  \theta\right)
\frac{d\theta^{\beta}}{d\xi}d\xi\text{.} \label{divergence}%
\end{equation}
From Eqs. (\ref{length}) and (\ref{divergence}), using the Cauchy-Schwarz
inequality, it follows that $\mathcal{I}\geq\mathcal{L}^{2}$. In particular,
the equality $\mathcal{I}=\mathcal{L}^{2}$ is obtained only when the integrand
in\ Eq. (\ref{divergence}) is constant along the path $\gamma_{\theta}$. The
thermodynamic length and the thermodynamic divergence can be regarded as
control measures of the dissipation of finite time thermodynamic processes. In
particular, while $\mathcal{I}$ measures the number of natural fluctuations
along a path, $\mathcal{L}$ is an indicator of the cumulative root-mean-square
deviations measured along the path \cite{crooks07}.

\subsection{Optimum paths}

Optimum paths $\gamma_{\theta}$ with $\theta\left(  \xi\right)  \overset
{\text{def}}{=}\left\{  \theta^{\alpha}\left(  \xi\right)  \right\}  $,
$1\leq\alpha\leq M$ where $M$ is the dimensionality of the parameter space,
and $0\leq\xi\leq\tau$ are paths characterized by the most favorable affine
time $\xi$ parametrization. Such a favorable time parametrization of the path
$\theta\left(  \xi\right)  $ yields the shortest thermodynamic length.
Specifically, minimization of the action functional represented by the length
$\mathcal{L}$ in\ Eq. (\ref{length}) by requiring its variation $\delta
\mathcal{L}$ equals zero yields, after some straightforward tensor algebra and
imposing that $\delta\theta^{\alpha}=0$ at the extremum, we obtain the
standard form for the geodesic equation,%
\begin{equation}
\frac{d^{2}\theta^{\alpha}}{d\xi^{2}}+\Gamma_{\beta\gamma}^{\alpha}%
\frac{d\theta^{\beta}}{d\xi}\frac{d\theta^{\gamma}}{d\xi}=0\text{.} \label{ge}%
\end{equation}
We note that the affine parameter $\xi$ is not unique since it is defined up
to changes of scale and origin. In summary, optimum paths are geodesic paths
$\theta^{\alpha}\left(  \xi\right)  $ that solve Eq. (\ref{ge}). The
quantities $\Gamma_{\nu\rho}^{\mu}$ in Eq. (\ref{ge}) are the Christoffel
connection coefficients of the second kind defined as \cite{felice90},
$\Gamma_{\beta\gamma}^{\alpha}\overset{\text{def}}{=}(1/2)g^{\alpha\delta
}\left(  \partial_{\beta}g_{\delta\gamma}+\partial_{\gamma}g_{\beta\delta
}-\partial_{\delta}g_{\beta\gamma}\right)  $ with $\partial_{\beta}%
\overset{\text{def}}{=}\partial/\partial\theta^{\beta}$. We point out that
optimum paths minimizing $\mathcal{L}\left(  \tau\right)  $ in\ Eq.
(\ref{length}) are also paths minimizing the divergence $\mathcal{I}\left(
\tau\right)  $ in\ Eq. (\ref{divergence}). Indeed, by minimizing
$\mathcal{I}\left(  \tau\right)  $ under the same working assumptions employed
in the minimization of the length, it can be shown\textbf{ }after some
straightforward computations that the optimum paths $\theta^{\alpha}\left(
\xi\right)  $ satisfy the equation%
\begin{equation}
\frac{d}{d\xi}\left[  \mathcal{F}_{\alpha\rho}\left(  \theta\right)
\frac{d\theta^{\alpha}}{d\xi}\right]  -\frac{1}{2}\frac{d\theta^{\alpha}}%
{d\xi}\frac{\partial\mathcal{F}_{\alpha\beta}\left(  \theta\right)  }%
{\partial\theta^{\rho}}\frac{d\theta^{\beta}}{d\xi}=0\text{.} \label{general1}%
\end{equation}
Interestingly, Eq. (\ref{general1}) is the information geometric analog of
Eqs. $(36)$ and $(6)$ in Refs. \cite{diosi96} and \cite{crooks17},
respectively. Since optimum paths are geodesic paths, the \textquotedblleft
thermodynamic\textquotedblright\ speed (henceforth, entropic speed)
$v_{\text{\textrm{E}}}$ defined as%
\begin{equation}
v_{\text{\textrm{E}}}\overset{\text{def}}{=}\sqrt{\frac{d\theta^{\alpha}}%
{d\xi}g_{\alpha\beta}\left(  \theta\right)  \frac{d\theta^{\beta}}{d\xi}%
}\text{,} \label{vthermo}%
\end{equation}
is constant when evaluated along these shortest paths. Furthermore, optimum
paths are also paths that correspond to constant entropy production rate
$r_{\text{\textrm{E}}}$, where $r_{\text{\textrm{E}}}$ is given by%
\begin{equation}
r_{\text{\textrm{E}}}\overset{\text{def}}{=}\frac{d\mathcal{I}}{d\tau}\text{,}
\label{EPR}%
\end{equation}
with $\mathcal{I}\left(  \tau\right)  $ defined in Eq. (\ref{divergence}) and
evaluated along the optimum paths. Finally, we point out that the entropy
production rate $r_{\text{\textrm{E}}}$ equals the squared invariant norm of
the speed $v_{\text{\textrm{E}}}$ when both quantities are evaluated along the
optimum paths. For the sake of forthcoming discussions, we shall be naming
lengths, divergences, and speeds as \textquotedblleft
entropic\textquotedblright\ quantities. Our proposed notion of efficiency is
inspired here by the definition of thermal efficiency of a heat engine
\cite{beretta05} and by the concept of efficiency of a quantum evolution in
the Riemannian approach to quantum mechanics as presented in Ref.
\cite{anandan90}.\ Specifically, replacing the condition of minimum energy
dispersion with the requirement of minimum entropy production, we find
it\textbf{ }convenient to define the entropic efficiency of an evolution along
a path of minimum entropic length joining the distinct initial and final
points on the information manifold as,%
\begin{equation}
\eta_{E}\overset{\text{def}}{=}1-\frac{r_{E}}{r}\text{,} \label{efficiency}%
\end{equation}
where $r\overset{\text{def}}{=}\max\left\{  \left\lceil r_{E}^{\left(
i\right)  }\right\rceil \right\}  $ is the maximum of the ceiling functions of
$r_{E}^{\left(  i\right)  }\in%
\mathbb{R}
_{+}\backslash\left\{  0\right\}  $ with the index \textquotedblleft%
$i$\textquotedblright\ labeling the distinct evolutions (of quantum mechanical
origin, in our case) being compared. The ceiling function maps $x\in%
\mathbb{R}
$ to the least integer greater than or equal to $x$. The quantity $r$ plays
the effective role of a normalizing factor that renders $\eta_{E}$
adimensional with $0\leq\eta_{E}\leq1$. Furthermore, in view of the fact that
we wish to rank the relative entropic performance of the various evolutions,
the quantity $r$ can be interpreted as the least integer upper bound of the
entropy production rate of the hottest among all cool paths available for each
evolution under consideration. Clearly, the entropic efficiency $\eta_{E}$ in
Eq. (\ref{efficiency}) assumes the ideal value $\eta_{E}=1$ when the evolution
is characterized by a path that is maximally cooled (that is, maximally
reversible). In such a case, the total entropy production remains ideally
constant during the evolution and, consequently, the rate of entropy
production $r_{E}$ approaches the limiting value of zero.

In the next section, we describe the quantum mechanical evolutions that
generate the probability paths that we use to present an information geometric
analysis of speed and minimum entropy production with some of the concepts we
have just introduced in the current section.

\section{The \textrm{su}$(2$; $%
\mathbb{C}
)$ Hamiltonian models}

In this section, we explain the manner in which the normalized pure states
that we consider emerge as outputs of suitable $\mathrm{su}\left(  2\text{; }%
\mathbb{C}
\right)  $ time-dependent Hamiltonian evolutions. These normalized pure states
are employed to describe, from a physics perspective, different types of
analog quantum search schemes \cite{alsing19,alsing19b}.

The quantum evolution that we consider is defined in terms of an Hamiltonian
operator $\mathcal{H}_{\mathrm{su}\left(  2\text{; }%
\mathbb{C}
\right)  }$ written as the most general linear superposition of the three
traceless and anti-Hermitian generators $\left\{  i\sigma_{x}\text{, }%
-i\sigma_{y}\text{, }i\sigma_{z}\right\}  $ of $\mathrm{su}\left(  2\text{; }%
\mathbb{C}
\right)  $, the Lie algebra of the special unitary group $\mathrm{SU}\left(
2\text{; }%
\mathbb{C}
\right)  $ \cite{sakurai94}, $\mathcal{H}_{\mathrm{su}\left(  2\text{; }%
\mathbb{C}
\right)  }\left(  t\right)  \overset{\text{def}}{=}a\left(  t\right)  \left(
i\sigma_{x}\right)  +b\left(  t\right)  \left(  -i\sigma_{y}\right)  +c\left(
t\right)  \left(  \text{ }i\sigma_{z}\right)  $. \ The quantities $a\left(
t\right)  $, $b\left(  t\right)  $, and $c\left(  t\right)  $ are
time-dependent complex coefficients while $\vec{\sigma}\overset{\text{def}}%
{=}\left(  \sigma_{x}\text{, }\sigma_{y}\text{, }\sigma_{z}\right)  $ is the
Pauli vector operator \cite{carlopra10,carlopra14}. In particular, by setting
$a\left(  t\right)  \overset{\text{def}}{=}-i\omega_{x}\left(  t\right)  $,
$b\left(  t\right)  \overset{\text{def}}{=}i\omega_{y}\left(  t\right)  $, and
$c\left(  t\right)  \overset{\text{def}}{=}-i\Omega\left(  t\right)  $, the
Hamiltonian $\mathcal{H}_{\mathrm{su}\left(  2\text{; }%
\mathbb{C}
\right)  }\left(  t\right)  $ becomes
\begin{equation}
\mathcal{H}_{\mathrm{su}\left(  2\text{; }%
\mathbb{C}
\right)  }\left(  t\right)  \overset{\text{def}}{=}\omega_{x}\left(  t\right)
\sigma_{x}+\omega_{y}\left(  t\right)  \sigma_{y}+\Omega\left(  t\right)
\sigma_{z}\text{.} \label{peter2}%
\end{equation}
In the language of $\mathrm{su}\left(  2\text{; }%
\mathbb{C}
\right)  $ Hamiltonian models, $\omega\left(  t\right)  \overset{\text{def}%
}{=}\omega_{x}\left(  t\right)  -i\omega_{y}\left(  t\right)  =\omega
_{\mathcal{H}}\left(  t\right)  e^{i\phi_{\omega}\left(  t\right)  }$ and
$\Omega\left(  t\right)  $ denote the so-called complex transverse field and
real longitudinal field, respectively. Clearly, $\omega_{\mathcal{H}}\left(
t\right)  $ denotes the modulus of $\omega\left(  t\right)  $. In what
follows, we specify that longitudinal fields $\Omega\left(  t\right)  $ are
oriented\textbf{\ }along the $z$-axis while transverse fields $\omega\left(
t\right)  $ lie in the $xy$-plane. Considering the quantum mechanical
evolution of a spin-$1/2$ particle (an electron, for instance) in an external
time-dependent magnetic field $\vec{B}\left(  t\right)  $, the Hamiltonian
$\mathcal{H}_{\mathrm{su}\left(  2\text{; }%
\mathbb{C}
\right)  }\left(  t\right)  $ in Eq. (\ref{peter2}) can be recast as
$\mathcal{H}_{\mathrm{su}\left(  2\text{; }%
\mathbb{C}
\right)  }\left(  t\right)  \overset{\text{def}}{=}-\vec{\mu}\cdot\vec
{B}\left(  t\right)  $, where $\vec{\mu}\overset{\text{def}}{=}\left(
e\hslash/2mc\right)  \vec{\sigma}$ is the magnetic moment of the electron with
$\mu_{\text{Bohr}}\overset{\text{def}}{=}e\hslash/(2mc)$ denoting the
so-called Bohr magneton. The quantity $m$ denotes the mass of an electron
while $\left\vert e\right\vert $ is the absolute value of the electric charge
of an electron. Furthermore, $c$ and $\hslash$ denote the speed of light and
the reduced Planck constant, respectively. The magnetic field $\vec{B}\left(
t\right)  $ can be decomposed as $\vec{B}\left(  t\right)  \overset
{\text{def}}{=}\vec{B}_{\perp}\left(  t\right)  +\vec{B}_{\parallel}\left(
t\right)  $, with $\vec{B}_{\perp}\left(  t\right)  \overset{\text{def}}%
{=}B_{x}\left(  t\right)  \hat{x}+B_{y}\left(  t\right)  \hat{y}$ and $\vec
{B}_{\parallel}\left(  t\right)  \overset{\text{def}}{=}B_{z}\left(  t\right)
\hat{z}$. It is straightforward to identify the link between the set of field
intensities $\left\{  \omega_{\mathcal{H}}\left(  t\right)  \text{, }%
\Omega_{\mathcal{H}}\left(  t\right)  \right\}  $ and the set of magnetic
field intensities $\left\{  B_{\perp}\left(  t\right)  \text{, }B_{\parallel
}\left(  t\right)  \right\}  $. In particular, we note that $B_{\perp}\left(
t\right)  \propto\omega_{\mathcal{H}}\left(  t\right)  $ and $B_{\parallel
}\left(  t\right)  \propto$ $\Omega_{\mathcal{H}}\left(  t\right)
\overset{\text{def}}{=}\left\vert \Omega\left(  t\right)  \right\vert $. In
terms of components, the exact relation between $\left\{  B_{x}\left(
t\right)  \text{, }B_{y}\left(  t\right)  \text{, }B_{z}\left(  t\right)
\right\}  $ and $\left\{  \omega_{x}\left(  t\right)  \text{, }\omega
_{y}\left(  t\right)  \text{, }\Omega\left(  t\right)  \right\}  $ is given by
$B_{x}\left(  t\right)  =-\left(  2mc/e\hslash\right)  \omega_{x}\left(
t\right)  $, $B_{y}\left(  t\right)  =-\left(  2mc/e\hslash\right)  \omega
_{y}\left(  t\right)  $, and $B_{z}\left(  t\right)  =-\left(  2mc/e\hslash
\right)  \Omega\left(  t\right)  $. Furthermore, in terms of field
intensities, we obtain $B_{\perp}\left(  t\right)  =\left(  2mc/\left\vert
e\right\vert \hslash\right)  \omega_{\mathcal{H}}\left(  t\right)  $, and
$B_{\parallel}\left(  t\right)  =\left(  2mc/\left\vert e\right\vert
\hslash\right)  \Omega_{\mathcal{H}}\left(  t\right)  $. Investigating the
quantum mechanical evolution of an electron specified by the Hamiltonian
$\mathcal{H}_{\mathrm{su}\left(  2\text{; }%
\mathbb{C}
\right)  }\left(  t\right)  $ in terms of exact analytical expressions of
complex probability amplitudes and/or real transition probabilities from an
initial source state to a final target state is a highly nontrivial matter.
The unitarity of the quantum mechanical evolution requires that the complex
probability amplitudes $\alpha\left(  t\right)  $ and $\beta\left(  t\right)
$ satisfy the normalization condition, $\left\vert \alpha\left(  t\right)
\right\vert ^{2}+\left\vert \beta\left(  t\right)  \right\vert ^{2}=1$. Given
the unitary evolution operator $\mathcal{U}\left(  t\right)  $ with
$i\hslash\mathcal{\dot{U}}\left(  t\right)  =\mathcal{H}_{\mathrm{su}\left(
2\text{; }%
\mathbb{C}
\right)  }\mathcal{U}\left(  t\right)  $ and $\mathcal{\dot{U}}\overset
{\text{def}}{=}\partial_{t}\mathcal{U}$, the temporal evolution of a quantum
source state $\left\vert s\right\rangle \overset{\text{def}}{=}x\left\vert
w\right\rangle +\sqrt{1-x^{2}}\left\vert w_{\perp}\right\rangle $ can be
specified by means of the mapping, $\left(  x\text{, }\sqrt{1-x^{2}}\right)
\rightarrow\left(  \alpha\left(  t\right)  x+\beta\left(  t\right)
\sqrt{1-x^{2}}\text{, }-\beta^{\ast}\left(  t\right)  x+\alpha^{\ast}\left(
t\right)  \sqrt{1-x^{2}}\right)  $ where $x\overset{\text{def}}{=}\left\langle
w|s\right\rangle $ is the quantum overlap. The set of orthonormal state
vectors\textbf{ }$\left\{  \left\vert w\right\rangle \text{, }\left\vert
w_{\perp}\right\rangle \right\}  $\textbf{ }span the two-dimensional
search\textbf{ }space\textbf{ }of the\textbf{ }$N=2^{n}$-dimensional complex
Hilbert space\textbf{ }$\mathcal{H}_{2}^{n}$. Therefore, the probability that
the source state $\left\vert s\right\rangle $ transitions into the target
state $\left\vert w\right\rangle $ under $\mathcal{U}\left(  t\right)  $ is
given by,%
\begin{equation}
\mathcal{P}_{\left\vert s\right\rangle \rightarrow\left\vert w\right\rangle
}\left(  t\right)  \overset{\text{def}}{=}\left\vert \left\langle
w|\mathcal{U}\left(  t\right)  |s\right\rangle \right\vert ^{2}=\left\vert
\alpha\left(  t\right)  \right\vert ^{2}x^{2}+\left\vert \beta\left(
t\right)  \right\vert ^{2}\left(  1-x^{2}\right)  +\left[  \alpha\left(
t\right)  \beta^{\ast}\left(  t\right)  +\alpha^{\ast}\left(  t\right)
\beta\left(  t\right)  \right]  x\sqrt{1-x^{2}}\text{.} \label{good}%
\end{equation}
It is clear from Eq. (\ref{good}) that, in order\textbf{\ }to compute the
exact analytical expression of transition probabilities, one needs to have the
exact analytical expression of the evolution operator $\mathcal{U}\left(
t\right)  $ in terms of the complex probability amplitudes $\alpha\left(
t\right)  $ and $\beta\left(  t\right)  $.

Following our previous work presented in Ref. \cite{alsing19} and exploiting
the results in Refs. \cite{messina14,grimaudo18}, we consider here four
quantum mechanical scenarios in which the transition probability
$\mathcal{P}_{\left\vert w_{\perp}\right\rangle \rightarrow\left\vert
w\right\rangle }\left(  t\right)  $ from an initial state $\left\vert
w_{\perp}\right\rangle $ to a final state $\left\vert w\right\rangle $, where
$\left\langle w_{\perp}|w\right\rangle =\delta_{w_{\perp}\text{, }w}$ and with
$\sigma_{z}\left\vert w\right\rangle =+\left\vert w\right\rangle $ and
$\sigma_{z}\left\vert w_{\perp}\right\rangle =-\left\vert w_{\perp
}\right\rangle $, can be expressed in an exact analytical manner. In all four
cases, we assume to be in a physical situation where $\dot{\phi}_{\omega
}\left(  t\right)  =\omega_{0}$, $\Omega\left(  t\right)  =-\frac{\hslash}%
{2}\omega_{0}$, and $\omega_{0}$ is a negative constant. We emphasize that,
from a formal mathematical viewpoint, more general temporal behaviors of
$\dot{\phi}_{\omega}\left(  t\right)  $ and $\Omega\left(  t\right)  $ could
have been chosen provided the so-called generalized Rabi condition as
presented in Ref. \cite{messina14,grimaudo18}, $\dot{\phi}_{\omega}\left(
t\right)  +(2/\hslash)\Omega\left(  t\right)  =0$, is satisfied. However, the
choice made appears to be more convenient from an experimental perspective.

The four scenarios can be formally distinguished by means of the temporal
expression of the field intensity $\omega_{\mathcal{H}}\left(  t\right)  $. In
the first case, we assume a constant field intensity $\omega_{\mathcal{H}%
}\left(  t\right)  $, $\omega_{\mathcal{H}}^{\left(  1\right)  }\left(
t\right)  \overset{\text{def}}{=}\Gamma$. This case defines the original Rabi
scenario where $\mathcal{P}_{\left\vert w_{\perp}\right\rangle \rightarrow
\left\vert w\right\rangle }\left(  t\right)  $ is given by, $\mathcal{P}%
_{\left\vert w_{\perp}\right\rangle \rightarrow\left\vert w\right\rangle
}^{\left(  1\right)  }\left(  t\right)  =\sin^{2}\left[  \left(
\Gamma/\hslash\right)  t\right]  $. In the remaining three cases, we consider
three generalized Rabi scenarios where the field intensity $\omega
_{\mathcal{H}}\left(  t\right)  $ exhibits oscillatory, power law decay, and
exponential law decay behaviors, $\omega_{\mathcal{H}}^{\left(  2\right)
}\left(  t\right)  \overset{\text{def}}{=}\Gamma\cos\left(  \lambda t\right)
$, $\omega_{\mathcal{H}}^{\left(  3\right)  }\left(  t\right)  \overset
{\text{def}}{=}\frac{\Gamma}{\left(  1+\lambda t\right)  ^{2}}$, and
$\omega_{\mathcal{H}}^{\left(  4\right)  }\left(  t\right)  \overset
{\text{def}}{=}\Gamma e^{-\lambda t}$, respectively. Observe that
$\omega_{\mathcal{H}}^{\left(  2\right)  }\left(  t\right)  $ is a positive
quantity on a temporal scale with $0\leq t\leq\left(  \pi/2\right)
\lambda^{-1}$. In all three cases, it can be shown that the transition
probability $\mathcal{P}_{\left\vert w_{\perp}\right\rangle \rightarrow
\left\vert w\right\rangle }^{\left(  j\right)  }\left(  t\right)  $ is given
by \cite{grimaudo18},%
\begin{equation}
\mathcal{P}_{\left\vert w_{\perp}\right\rangle \rightarrow\left\vert
w\right\rangle }^{\left(  j\right)  }\left(  t\right)  =\sin^{2}\left[
\int_{0}^{t}\frac{\omega_{\mathcal{H}}^{\left(  j\right)  }\left(  t^{\prime
}\right)  }{\hslash}dt^{\prime}\right]  \text{,} \label{tp2}%
\end{equation}
for any $j\in\left\{  2\text{, }3\text{, }4\right\}  $. Interestingly, being
on resonance, the transition probability in all four cases depends only on the
integral of the transverse field intensity $\omega_{\mathcal{H}}\left(
t\right)  $. In this paper, the chosen expressions of $\omega_{\mathcal{H}%
}\left(  t\right)  $ serve to specify the particular type of behavior of the
analog quantum search algorithms that correspond to the time-dependent
two-level quantum systems, the latter being characterized by the selected
transverse field intensity.

A summary of the main properties of the four quantum evolutions that we
consider in this paper appear in Table II. The transition probabilities
$\mathcal{P}_{\left\vert w_{\perp}\right\rangle \rightarrow\left\vert
w\right\rangle }^{\left(  k\right)  }\left(  t\right)  $ with $1\leq k\leq4$
are the essential quantities that we use to construct our parametrized output
quantum states in the next section.

\begin{table}[t]
\centering
\begin{tabular}
[c]{c|c|c|c}\hline\hline
Rabi Scenario & Tranversal Magnetic Field Intensity, $B_{\perp}\left(
t\right)  $ & Resonance Condition & Complex Tranverse Field, $\omega\left(
t\right)  $\\\hline
original & $\frac{2mc}{\left\vert e\right\vert \hslash}\Gamma$ &
$B_{\parallel}=\frac{mc}{e}\omega_{0}$ & $\Gamma e^{i\phi_{\omega}\left(
t\right)  }$\\
generalized & $\frac{2mc}{\left\vert e\right\vert \hslash}\omega_{\mathcal{H}%
}\left(  t\right)  $ & $B_{\parallel}\left(  t\right)  =\frac{mc}{e}\dot{\phi
}_{\omega}\left(  t\right)  $ & $\Gamma\cos\left(  \lambda t\right)
e^{i\phi_{\omega}\left(  t\right)  }$\\
generalized & $\frac{2mc}{\left\vert e\right\vert \hslash}\omega_{\mathcal{H}%
}\left(  t\right)  $ & $B_{\parallel}\left(  t\right)  =\frac{mc}{e}\dot{\phi
}_{\omega}\left(  t\right)  $ & $\frac{\Gamma}{\left(  1+\lambda t\right)
^{2}}e^{i\phi_{\omega}\left(  t\right)  }$\\
generalized & $\frac{2mc}{\left\vert e\right\vert \hslash}\omega_{\mathcal{H}%
}\left(  t\right)  $ & $B_{\parallel}\left(  t\right)  =\frac{mc}{e}\dot{\phi
}_{\omega}\left(  t\right)  $ & $\Gamma e^{-\lambda t}e^{i\phi_{\omega}\left(
t\right)  }$\\\hline
\end{tabular}
\caption{Schematic description of the complex transverse field $\omega\left(
t\right)  $, the resonance condition in terms of the longitudinal magnetic
field intensity $B_{\parallel}\left(  t\right)  $, and the transverse magnetic
field intensity $B_{\perp}\left(  t\right)  $ in the chosen four quantum
mechanical Rabi scenarios.}%
\end{table}

\section{Optimum paths, entropic speed, and entropy production rate}

In this section, we first introduce the Riemannian metrization on the
parameter manifold specified by the Fisher information (see Eq. (\ref{JIJ2}))
introduced in Sec. II evaluated along the parametrized squared probability
amplitudes obtained from the analysis of the temporal evolution of a
spin-$1/2$ particle in an external time-dependent magnetic field that
characterizes the $\mathrm{su}\left(  2\text{; }%
\mathbb{C}
\right)  $ Hamiltonian model described in\ Sec. III. Then, using a minimum
action method to transfer a quantum system from an initial state to a final
state on the manifold in finite time, we show that the minimizing (optimum)
path is the shortest (geodesic) path between the two states, and, in
particular, minimizes also the total entropy production that occurs during the
transfer. Finally, by calculating the entropic speed (see Eq. (\ref{vthermo}))
and the entropy production rate (see Eq. (\ref{EPR})) along the optimum
transfer paths in the physical scenarios outlined in Sec. III and Table II, we
verify in a transparent quantitative manner that to a faster transfer there
corresponds necessarily a higher entropy production rate.

\subsection{From quantum evolutions to probability paths}

In our work, the parameter $\theta$ denotes the statistical version of the
elapsed time $t$ where we assume that $\theta$ is an experimental parameter
that can be determined by measurement of a conventional observable that varies
with time (for instance, the transverse magnetic field intensity\textbf{
}$B_{\perp}\left(  t\right)  =\left(  2mc/\left\vert e\right\vert
\hslash\right)  \omega_{\mathcal{H}}\left(  t\right)  $. We point out that
choosing the elapsed time as the experimentally controllable statistical
parameter is not unusual. For example, the comparison between the inverse
temperature $\beta\overset{\text{def}}{=}\left(  k_{\text{B}}T\right)  ^{-1}$
with $k_{\text{B}}$ denoting the Boltzmann constant and $\theta/\hslash$ is
reminiscent of the well-known connection between statistical mechanics and
quantum mechanics in terms of the so-called Wick rotation. Specifically, by
replacing $\beta$ with the imaginary time $it/\hslash$, the Wick rotation
allows one to compute quantum mechanical probability amplitudes just as one
calculates averages of observables in statistical mechanics. Therefore, by
minimizing the entropy production, we shall find the optimum paths on the
manifold of state space parametrized by $\theta$ along which one drives the
system. To be more specific, in view of the connection between analog quantum
search and two-level quantum systems \cite{alsing19,alsing19b}, we assume that
the output of a continuous-time quantum search algorithm where the input is
the normalized $N=2^{n}$-dimensional $n$-qubit source state $\left\vert
s\right\rangle \overset{\text{def}}{=}\left\vert \psi\left(  \theta
_{0}\right)  \right\rangle $ can be described as, $\left\vert \psi\left(
\theta\right)  \right\rangle \overset{\text{def}}{=}e^{i\varphi_{w}\left(
\theta\right)  }\sqrt{p_{w}\left(  \theta\right)  }\left\vert w\right\rangle
+e^{i\varphi_{w_{\perp}}\left(  \theta\right)  }\sqrt{p_{w_{\perp}}\left(
\theta\right)  }\left\vert w_{\perp}\right\rangle $. The $N$-dimensional
normalized output state $\left\vert \psi\left(  \theta\right)  \right\rangle $
belongs to the two-dimensional subspace of the $n$-qubit complex Hilbert space
$\mathcal{H}_{2}^{n}$ spanned by the set of orthonormal state vectors
$\left\{  \left\vert w\right\rangle \text{, }\left\vert w_{\perp}\right\rangle
\right\}  $ and containing the source state $\left\vert s\right\rangle $. The
squared probability amplitude $p_{w}\left(  \theta\right)  \overset
{\text{def}}{=}\left\vert \left\langle w|\psi\left(  \theta\right)
\right\rangle \right\vert ^{2}$ and $p_{w_{\perp}}\left(  \theta\right)
\overset{\text{def}}{=}\left\vert \left\langle w_{\perp}|\psi\left(
\theta\right)  \right\rangle \right\vert ^{2}$ denote the success and failure
probabilities of the search algorithm, respectively. Finally, $\varphi
_{w}\left(  \theta\right)  $ and $\varphi_{w_{\perp}}\left(  \theta\right)  $
are real quantum phases of the states $\left\vert w\right\rangle $ and
$\left\vert w_{\perp}\right\rangle $, respectively. The quantum state
$\left\vert \psi\left(  \theta\right)  \right\rangle $ is parametrized in
terms of a single continuous real parameter that emerges from the (computing)
elapsed time of the algorithm. As briefly mentioned earlier, this parameter
$\theta$ plays the role of a statistical macrovariable used to distinguish
neighboring quantum states $\left\vert \psi\left(  \theta\right)
\right\rangle $ and $\left\vert \psi\left(  \theta\right)  \right\rangle
+\left\vert d\psi\left(  \theta\right)  \right\rangle $ along a path through
the space of quantum mechanical pure states. In summary, given our working
conditions outlined earlier and assuming to have $\left\vert \psi\left(
\theta_{0}\right)  \right\rangle $ $=\left\vert w_{\perp}\right\rangle $ as
our input state, we shall essentially focus on the space of probability
distributions $\left\{  p\left(  \theta\right)  \right\}  $ with $p\left(
\theta\right)  \overset{\text{def}}{=}\left(  p_{w}\left(  \theta\right)
\text{, }p_{w_{\perp}}\left(  \theta\right)  \right)  $ and with the natural
Riemannian distinguishability metric given by the Fisher information metric in
Eq. (\ref{JIJ2}) (which, under suitably chosen working conditions
\cite{caves94}, can be taken proportional to the Fubini-Study metric),
$\left\vert \psi\left(  \theta\right)  \right\rangle \mapsto p\left(
\theta\right)  =\left(  p_{w}\left(  \theta\right)  \text{, }p_{w_{\perp}%
}\left(  \theta\right)  \right)  =\left(  \left\vert \left\langle
w|\psi\left(  \theta\right)  \right\rangle \right\vert ^{2}\text{, }\left\vert
\left\langle w_{\perp}|\psi\left(  \theta\right)  \right\rangle \right\vert
^{2}\right)  $. In Fig.\textbf{ }$1$\textbf{, }we report the behavior of the
Fisher information evaluated along the probabilities obtained from an\textbf{
}$\mathrm{su}\left(  2\text{; }%
\mathbb{C}
\right)  $\textbf{ }quantum evolution specified by an exponentially decaying
transverse field intensity\textbf{ }$\omega_{\mathcal{H}}^{\left(  4\right)
}\left(  t\right)  $.

\begin{figure}[t]
\centering
\includegraphics[width=1\textwidth] {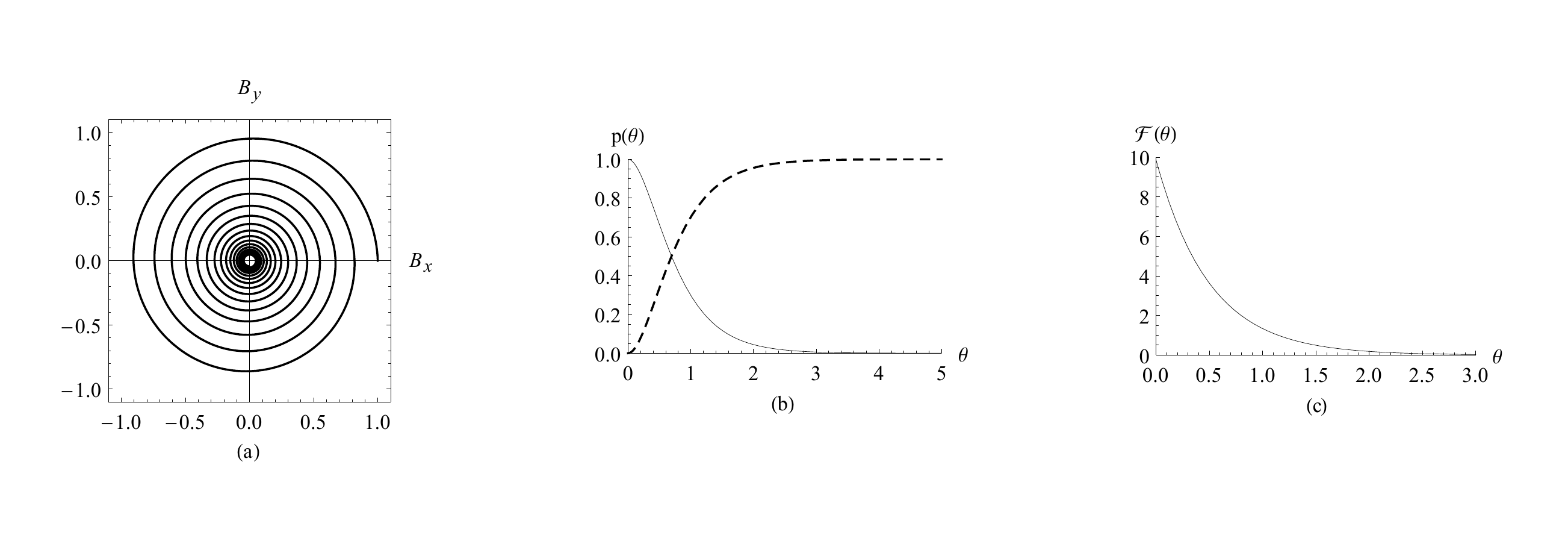}\caption{Illustrative depiction of
the behavior of the Fisher information $\mathcal{F}\left(  \theta\right)  $
versus $\theta$ (plot (c)) defined in terms of the transition probabilities
$p_{w}\left(  \theta\right)  $ (dashed line) and $p_{w_{\bot}}\left(
\theta\right)  $ (solid line) (plot (b)). These probabilities, in turn, emerge
from the specific external magnetic field configuration $\vec{B}=\vec{B}%
_{\bot}+\vec{B}_{\parallel}$ with $\vec{B}_{\bot}=B_{x}\hat{x}+B_{y}\hat{y}$
(plot (a)) that characterizes the $\mathrm{su}\left(  2\text{; }%
\mathbb{C}\right)  $ Hamiltonian model being considered. In plots (b) and (c),
we set $\Gamma/\hslash\lambda=\pi/2$ and $\lambda=1$. In plot (a), the
two-dimensional parametric plot of the transverse magnetic field components,
we set $\left\vert \omega_{0}\right\vert =10\pi$ and $\lambda=1$. Furthermore,
for simplifying normalization purposes, we also set $\Gamma/\mu_{\text{Bohr}%
}=1$ with $\mu_{\text{Bohr}}$ denoting the Bohr magneton in plot (a). All
physical quantities are assumed to be suitably expressed in terms of the MKSA
unit system. Finally, the $\mathrm{su}\left(  2\text{; }\mathbb{C}\right)  $
Hamiltonian model considered in this figure corresponds to the fourth scenario
studied in the paper.}%
\end{figure}

\subsection{Illustrative examples}

In order to compute the entropic speed $v_{\mathrm{E}}$ and the entropy
production rate $r_{\mathrm{E}}$ along the optimum cooling paths, we clearly
need to first locate such paths. The expression of paths $\gamma_{\theta}$
with $\theta=\theta\left(  \xi\right)  $ and $\xi$ being an affine parameter
depends upon the specific parametric behavior of the Fisher information
$\mathcal{F}\left(  \theta\right)  =E_{\theta}\left[  \left\{  \partial
_{\theta}\log\left[  p_{x}\left(  \theta\right)  \right]  \right\}
^{2}\right]  $ evaluated along the probability paths $p\left(  \theta\right)
\overset{\text{def}}{=}\left(  p_{w}\left(  \theta\right)  \text{,
}p_{w_{\perp}}\left(  \theta\right)  \right)  $. The probabilities
$p_{w}\left(  \theta\right)  $ and $p_{w_{\perp}}\left(  \theta\right)  $
emerge, in turn, from analyzing the chosen Schr\"{o}dinger evolutions
specified in the previous subsection. Therefore, the Fisher information plays
a key role in our proposed information geometric analysis of quantum
mechanical evolutions.

\subsubsection{Constant Fisher information}

In the first quantum mechanical scenario that we consider, the space of
probability distributions $\left\{  p\left(  \theta\right)  \right\}  $ with
$p\left(  \theta\right)  \overset{\text{def}}{=}\left(  p_{w}\left(
\theta\right)  \text{, }p_{w_{\perp}}\left(  \theta\right)  \right)  $ is
defined in terms of the success and failure probabilities%
\begin{equation}
p_{w}\left(  \theta\right)  \overset{\text{def}}{=}\sin^{2}\left(
\frac{\Gamma}{\hslash}\theta\right)  \text{, and }p_{w_{\perp}}\left(
\theta\right)  \overset{\text{def}}{=}\cos^{2}\left(  \frac{\Gamma}{\hslash
}\theta\right)  \text{,} \label{1a}%
\end{equation}
respectively. We note that the probabilities in Eq. (\ref{1a}) exhibit a
periodic oscillatory behavior with period given by $T\overset{\text{def}}%
{=}\left(  \pi\hslash\right)  /\Gamma$. Using Eq. (\ref{1a}), the Fisher
information $\mathcal{F}\left(  \theta\right)  $ in Eq. (\ref{JIJ2}) becomes a
constant quantity $\mathcal{F}_{0}$ with $\mathcal{F}\left(  \theta\right)
=\mathcal{F}_{0}\overset{\text{def}}{=}4\left(  \Gamma/\hslash\right)  ^{2}$.
Furthermore, since we are focusing on a single parameter probability
path\textbf{ }$\gamma_{\theta}$\textbf{ }with\textbf{ }$\theta=\theta\left(
\xi\right)  $, the geodesic equations in Eq.(\ref{general1}) reduce to,%
\begin{equation}
\frac{d^{2}\theta}{d\xi^{2}}+\frac{1}{2\mathcal{F}}\frac{d\mathcal{F}}%
{d\theta}\left(  \frac{d\theta}{d\xi}\right)  ^{2}=0\text{.} \label{odeti}%
\end{equation}
In particular, given that $\mathcal{F}\left(  \theta\right)  =\mathcal{F}_{0}%
$, Eq. (\ref{odeti}) yields, $d^{2}\theta/d\xi^{2}$ $=0$. Assuming non
vanishing positive initial conditions $\theta\left(  \xi_{0}\right)
=\theta_{0}$ and $\dot{\theta}\left(  \xi_{0}\right)  =\dot{\theta}_{0}$,
integration of the geodesic equation is trivial and leads to the following
optimum paths,%
\begin{equation}
\theta\left(  \xi\right)  =\theta_{0}+\dot{\theta}_{0}\left(  \xi-\xi
_{0}\right)  \text{.} \label{geo1a}%
\end{equation}
From the knowledge of the optimum paths, we can finally compute both the
entropic speed $v_{\mathrm{E}}$ and the entropy production rate $r_{\mathrm{E}%
}$ that characterize the geodesic motion on the statistical manifold being
considered. Specifically, evaluating the entropic speed $v_{\mathrm{E}}$ in
Eq. (\ref{vthermo}) and the total entropy production $r_{\mathrm{E}}$ in Eq.
(\ref{divergence}) along the optimum paths in\ Eq. (\ref{geo1a}), we obtain%
\begin{equation}
v_{\mathrm{E}}\left(  \Gamma\right)  =\frac{\Gamma}{\hslash}\dot{\theta}%
_{0}\text{, and }r_{\mathrm{E}}\left(  \Gamma\right)  =\left(  \frac{\Gamma
}{\hslash}\right)  ^{2}\dot{\theta}_{0}^{2}\text{,} \label{speed1}%
\end{equation}
respectively. From Eq. (\ref{speed1}), we note that $v_{\mathrm{E}}\left(
\Gamma\right)  \propto\Gamma$ while $r_{\mathrm{E}}\left(  \Gamma\right)
\propto\Gamma^{2}$. Therefore, the magnitude $\omega_{\mathcal{H}}^{\left(
1\right)  }\left(  t\right)  =\Gamma$ of the complex transverse field that
specifies the $\mathrm{su}\left(  2\text{; }%
\mathbb{C}
\right)  $ Hamiltonian in Eq. (\ref{peter2}) is the essential quantity that
one needs to manipulate in order to find a convenient tradeoff between speed
and entropy production rate within our information geometric analysis of
quantum mechanical evolutions.

\subsubsection{Oscillatory behavior of the Fisher information}

In the second quantum mechanical scenario that we analyze, the space of
probability distributions $\left\{  p\left(  \theta\right)  \right\}  $ with
$p\left(  \theta\right)  \overset{\text{def}}{=}\left(  p_{w}\left(
\theta\right)  \text{, }p_{w_{\perp}}\left(  \theta\right)  \right)  $ is
defined in terms of the success and failure probabilities%
\begin{equation}
p_{w}\left(  \theta\right)  \overset{\text{def}}{=}\sin^{2}\left[
\frac{\Gamma}{\hslash\lambda}\sin\left(  \lambda\theta\right)  \right]
\text{, and }p_{w_{\perp}}\left(  \theta\right)  \overset{\text{def}}{=}%
\cos^{2}\left[  \frac{\Gamma}{\hslash\lambda}\sin\left(  \lambda\theta\right)
\right]  \text{,} \label{5b}%
\end{equation}
respectively. From Eq. (\ref{5b}), we observe that the probabilities
$p_{w}\left(  \theta\right)  $ and $p_{w_{\perp}}\left(  \theta\right)  $
exhibit a periodic oscillatory behavior with period given by $T\overset
{\text{def}}{=}\pi/\lambda$. In particular, $p_{w}\left(  \theta\right)  $
reaches its maximum value $\sin^{2}\left[  \Gamma/\left(  \hslash
\lambda\right)  \right]  $ at $t^{\ast}\overset{\text{def}}{=}\pi/\left(
2\lambda\right)  $. Therefore, in order for $p_{w}\left(  \theta\right)  $ to
reach a maximum value equal to one, we need to impose the constraint
$\Gamma=(h/4)\lambda$. Employing Eq. (\ref{5b}), the Fisher information
$\mathcal{F}\left(  \theta\right)  $ becomes, $\mathcal{F}\left(
\theta\right)  =4\left(  \Gamma/\hslash\right)  ^{2}\cos^{2}\left(
\lambda\theta\right)  $. In the case being considered, Eq. (\ref{odeti})
becomes $d^{2}\theta/d\xi^{2}-\lambda\tan\left(  \lambda\theta\right)  \left(
d\theta/d\xi\right)  ^{2}=0$. Assuming nonvanishing positive initial
conditions $\theta\left(  \xi_{0}\right)  =\theta_{0}$ and $\dot{\theta
}\left(  \xi_{0}\right)  =\dot{\theta}_{0}$, integration of the geodesic
equation yields optimum paths $\theta\left(  \xi\right)  $ whose general
expression can be recast as,
\begin{equation}
\theta\left(  \xi\right)  =\theta_{0}+\frac{\sqrt{1-\lambda^{2}\xi_{0}^{2}}%
}{\lambda}\dot{\theta}_{0}\left[  \sin^{-1}\left(  \lambda\xi\right)
-\sin^{-1}\left(  \lambda\xi_{0}\right)  \right]  \text{.} \label{geo2}%
\end{equation}
From the knowledge of the optimum paths, we can find both the entropic speed
$v_{\mathrm{E}}$ and the entropy production rate $r_{\mathrm{E}}$ that specify
the geodesic motion on the statistical manifold being considered.
Specifically, evaluating the entropic speed $v_{\mathrm{E}}$ in Eq.
(\ref{vthermo}) and the total entropy production $r_{\mathrm{E}}$ in Eq.
(\ref{divergence}) along the optimum paths in\ Eq. (\ref{geo2}), we find%
\begin{equation}
v_{\mathrm{E}}\left(  \Gamma\right)  =\frac{\Gamma}{\hslash}\left\vert
\cos\left(  \lambda\theta_{0}\right)  \right\vert \dot{\theta}_{0}\text{, and
}r_{\mathrm{E}}\left(  \Gamma\right)  =\left(  \frac{\Gamma}{\hslash}\right)
^{2}\cos^{2}\left(  \lambda\theta_{0}\right)  \dot{\theta}_{0}^{2}\text{,}
\label{speed1b}%
\end{equation}
respectively. Comparing Eqs. (\ref{speed1b}) and (\ref{speed1}), we note that
with respect to the first scenario, this second scenario is characterized by a
geodesic motion generating cooler optimum paths explored with a smaller
entropic speed.

\begin{figure}[t]
\centering
\includegraphics[width=1\textwidth] {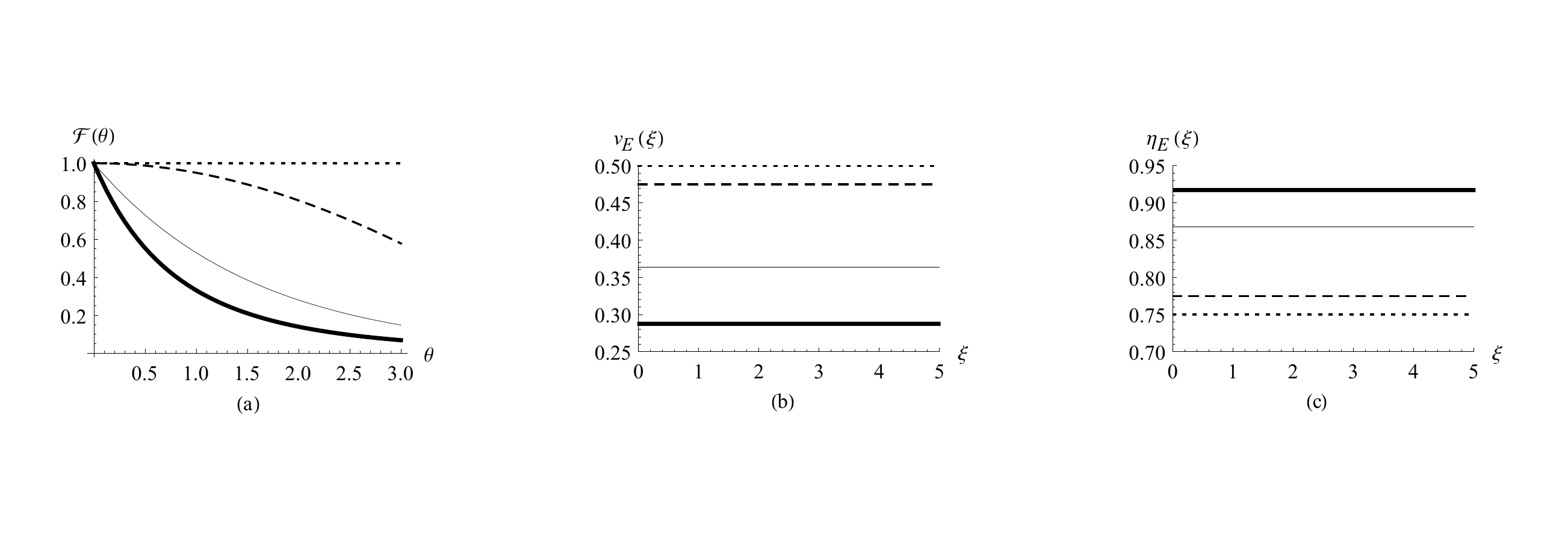}\caption{Illustrative depiction of
the behavior of the Fisher information $\mathcal{F}\left(  \theta\right)  $
versus $\theta$ (plot (a)), the entropic speed $v_{E}\left(  \xi\right)  $
versus $\xi$ (plot (b)), and the entropic efficiency $\eta_{E}\left(
\xi\right)  $ versus $\xi$ (plot (c)). Dotted, dashed, thin solid, and thick
solid lines in plots (a), (b), and (c) correspond to constant, oscillatory,
exponential decay, and power law decay of the Fisher information,
respectively. In all plots, we impose $\Gamma/(\hslash\lambda)=\pi/2$,
$\lambda=1/\pi$, $\theta_{0}=1$, and $\dot{\theta}_{0}=1$. Finally, all
physical quantities are assumed to be suitably expressed in terms of the MKSA
unit system.}%
\end{figure}

\subsubsection{Power law decay of the Fisher information}

In the third quantum mechanical scenario that we investigate, the space of
probability distributions $\left\{  p\left(  \theta\right)  \right\}  $ with
$p\left(  \theta\right)  \overset{\text{def}}{=}\left(  p_{w}\left(
\theta\right)  \text{, }p_{w_{\perp}}\left(  \theta\right)  \right)  $ is
specified by means of the success and failure probabilities%
\begin{equation}
p_{w}\left(  \theta\right)  \overset{\text{def}}{=}\sin^{2}\left[
\frac{\Gamma}{\hslash\lambda}\left(  1-\frac{1}{1+\lambda\theta}\right)
\right]  \text{, and }p_{w_{\perp}}\left(  \theta\right)  \overset{\text{def}%
}{=}\cos^{2}\left[  \frac{\Gamma}{\hslash\lambda}\left(  1-\frac{1}%
{1+\lambda\theta}\right)  \right]  \text{,} \label{4b}%
\end{equation}
respectively. The probability $p_{w}\left(  \theta\right)  $ in Eq. (\ref{4b})
exhibits an asymptotic monotonic convergence to one provided that
$\Gamma=(h/4)\lambda$. Furthermore, making use of Eq. (\ref{4b}), the Fisher
information $\mathcal{F}\left(  \theta\right)  $ becomes, $\mathcal{F}\left(
\theta\right)  =4\left(  \Gamma/\hslash\right)  ^{2}\left(  1+\lambda
\theta\right)  ^{-4}$. In this case, Eq. (\ref{odeti}) can be expressed as
$d^{2}\theta/d\xi^{2}-\left[  2\lambda/\left(  1+\lambda\theta\right)
\right]  \left(  d\theta/d\xi\right)  ^{2}=0$. Assuming nonvanishing positive
initial conditions $\theta\left(  \xi_{0}\right)  =\theta_{0}$ and
$\dot{\theta}\left(  \xi_{0}\right)  =\dot{\theta}_{0}$ and integrating this
geodesic equation , the optimum paths become%
\begin{equation}
\theta\left(  \xi\right)  =\frac{\left(  1+\lambda\theta_{0}\right)
^{2}+\lambda\dot{\theta}_{0}\left[  \left(  \xi-\xi_{0}\right)  -\frac
{1+\lambda\theta_{0}}{\lambda\dot{\theta}_{0}}\right]  }{\lambda^{2}%
\dot{\theta}_{0}\left[  \frac{1+\lambda\theta_{0}}{\lambda\dot{\theta}_{0}%
}-\left(  \xi-\xi_{0}\right)  \right]  }\text{.} \label{geo3}%
\end{equation}
Once again, from the knowledge of the optimum paths, we can calculate both the
entropic speed and the entropy production rate that characterize the geodesic
motion on the statistical manifold being considered. Specifically, evaluating
the entropic speed $v_{\mathrm{E}}$ in Eq. (\ref{vthermo}) and the entropy
production rate $r_{\mathrm{E}}$ in Eq.(\ref{EPR}) along the optimum paths
in\ Eq. (\ref{geo3}), we obtain%
\begin{equation}
v_{\mathrm{E}}\left(  \Gamma\right)  =\frac{\Gamma}{\hslash}\frac{1}{\left[
1+\lambda\left(  \Gamma\right)  \theta_{0}\right]  ^{2}}\dot{\theta}%
_{0}\text{, and }r_{\mathrm{E}}\left(  \Gamma\right)  =\left(  \frac{\Gamma
}{\hslash}\right)  ^{2}\frac{1}{\left[  1+\lambda\left(  \Gamma\right)
\theta_{0}\right]  ^{4}}\dot{\theta}_{0}^{2}\text{,} \label{speed3}%
\end{equation}
respectively, where $\lambda\left(  \Gamma\right)  \overset{\text{def}}%
{=}\left(  4\Gamma\right)  /h$. \ In analogy to the first and second
scenarios, the motion on the manifold associated with the third scenario
proceeds at constant entropic speed $v_{\mathrm{E}}$ and, thus, exhibits
minimum entropy production. In particular, this third scenario is
characterized by a geodesic motion that yields optimum paths cooler than those
found in the second scenario.

\subsubsection{Exponential decay of the Fisher information}

\begin{figure}[t]
\centering
\includegraphics[width=0.35\textwidth] {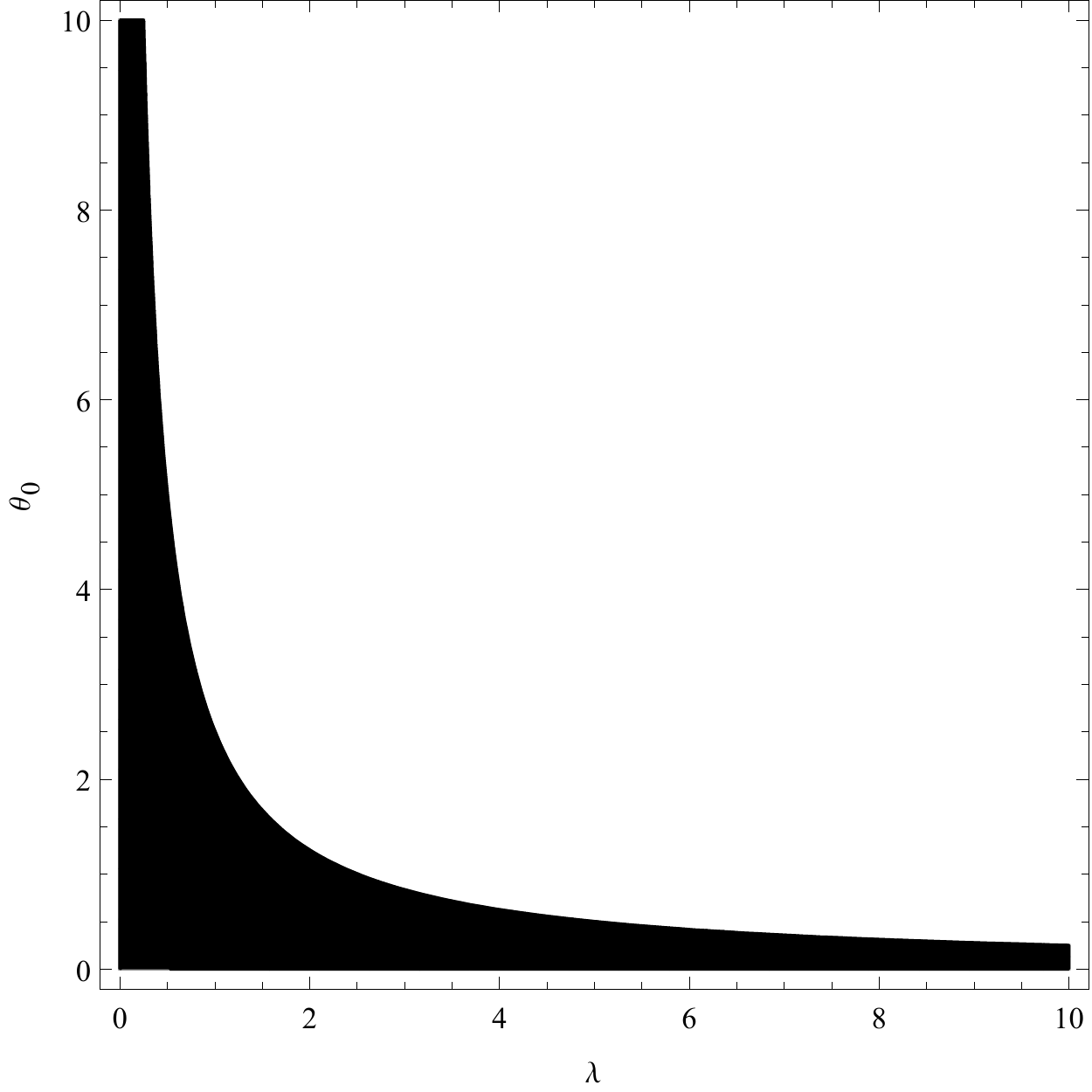}\caption{Plot of the
two-dimensional parametric region $\mathcal{P}$ where the exponential decay
strategy outperforms the power law decay strategy in terms of higher entropic
speed values. The black region $\mathcal{P}$ tends to vanish as the values of
$\lambda$ become sufficiently large.}%
\label{fig4}%
\end{figure}

In the fourth and last quantum mechanical scenario that we analyze, the space
of probability distributions $\left\{  p\left(  \theta\right)  \right\}  $
with $p\left(  \theta\right)  \overset{\text{def}}{=}\left(  p_{w}\left(
\theta\right)  \text{, }p_{w_{\perp}}\left(  \theta\right)  \right)  $ is
defined in terms of the success and failure probabilities%
\begin{equation}
p_{w}\left(  \theta\right)  \overset{\text{def}}{=}\sin^{2}\left[
\frac{\Gamma}{\hslash\lambda}\left(  1-e^{-\lambda\theta}\right)  \right]
\text{, and }p_{w_{\perp}}\left(  \theta\right)  \overset{\text{def}}{=}%
\cos^{2}\left[  \frac{\Gamma}{\hslash\lambda}\left(  1-e^{-\lambda\theta
}\right)  \right]  \text{,} \label{3a}%
\end{equation}
respectively. The probability $p_{w}\left(  \theta\right)  $ in Eq. (\ref{3a})
exhibits an asymptotic monotonic convergence to one provided that
$\Gamma=(h/4)\lambda$. Moreover, using Eq. (\ref{3a}), the Fisher information
$\mathcal{F}\left(  \theta\right)  $ becomes $\mathcal{F}\left(
\theta\right)  =4\left(  \Gamma/\hslash\right)  ^{2}e^{-2\lambda\theta}$. In
this case, the geodesic equation becomes $d^{2}\theta/d\xi^{2}-\lambda\left(
d\theta/d\xi\right)  ^{2}=0$. Considering nonvanishing positive initial
conditions $\theta\left(  \xi_{0}\right)  =\theta_{0}$ and $\dot{\theta
}\left(  \xi_{0}\right)  =\dot{\theta}_{0}$, integration of the geodesic
equation yields the following optimum paths,%
\begin{equation}
\theta\left(  \xi\right)  =\theta_{0}-\frac{1}{\lambda}\log\left[
1-\lambda\dot{\theta}_{0}\left(  \xi-\xi_{0}\right)  \right]  \text{.}
\label{geo4}%
\end{equation}
As pointed out in the previous illustrative examples, from the knowledge of
the optimum paths, we can determine both the entropic speed $v_{\mathrm{E}}$
and the entropy production rate $r_{\mathrm{E}}$ that characterize the
geodesic motion on the statistical manifold under consideration. Specifically,
evaluating the entropic speed $v_{\mathrm{E}}$ in Eq. (\ref{vthermo}) and the
total entropy production in Eq. (\ref{divergence}) along the optimum paths
in\ Eq. (\ref{geo4}), we find%
\begin{equation}
v_{\mathrm{E}}\left(  \Gamma\right)  =\frac{\Gamma}{\hslash}e^{-\lambda\left(
\Gamma\right)  \theta_{0}}\dot{\theta}_{0}\text{, and }r_{\mathrm{E}}\left(
\Gamma\right)  =\left(  \frac{\Gamma}{\hslash}\right)  ^{2}e^{-2\lambda\left(
\Gamma\right)  \theta_{0}}\dot{\theta}_{0}^{2}\text{,} \label{speed4}%
\end{equation}
respectively, where $\lambda\left(  \Gamma\right)  \overset{\text{def}}%
{=}\left(  4\Gamma\right)  /h$. From the comparison of Eqs. (\ref{speed3}) and
(\ref{speed4}), we conclude that this last scenario can exhibit the coolest
optimum paths, albeit with the path exploration occurring at the slowest
entropic speed for values of $\lambda\left(  \Gamma\right)  $ sufficiently
large. More specifically, we observe that $0\leq e^{-\lambda\theta_{0}}%
\leq1/\left(  1+\lambda\theta_{0}\right)  ^{2}\leq\left\vert \cos\left(
\lambda\theta_{0}\right)  \right\vert \leq1$ when $\theta_{0}\in%
\mathbb{R}
_{+}$ and $\lambda\left(  \Gamma\right)  \overset{\text{def}}{=}\left(
4\Gamma\right)  /h\gg1$. However, when $0\leq\lambda\lesssim1$, the
exponential-decay strategy can outperform the power-law strategy in terms of
entropic speed. In Fig. $2$ we plot the constant values of the entropic
speed\textbf{ }$v_{\mathrm{E}}$\textbf{ }and the entropic efficiency\textbf{
}$\eta_{\mathrm{E}}$\textbf{ }that emerge from the particular parametric
expression of $\mathcal{F}\left(  \theta\right)  $. The expression of the
Fisher information, in turn, depends on the particular $\mathrm{su}\left(
2\text{; }%
\mathbb{C}
\right)  $\textbf{ }Hamiltonian model considered. More formally, combining
Eqs. (\ref{speed3}) and (\ref{speed4}), we get $v_{\mathrm{E}}^{\left(
\text{power-law}\right)  }\left(  \Gamma\right)  =\left[  e^{\lambda\theta
_{0}}/\left(  1+\lambda\theta_{0}\right)  ^{2}\right]  v_{\mathrm{E}}^{\left(
\text{power-law}\right)  }\left(  \Gamma\right)  $. Introducing the function
$f_{\mathcal{P}}\left(  \lambda\text{, }\theta_{0}\right)  \overset
{\text{def}}{=}$\textbf{ }$e^{\lambda\theta_{0}}/\left(  1+\lambda\theta
_{0}\right)  ^{2}$, the two-dimensional parametric region\textbf{
}$\mathcal{P}$ where the exponential law decay strategy yields entropic speed
values higher than those of the power law decay strategy is given by,
$\mathcal{P}\overset{\text{def}}{=}\left\{  \left(  \lambda\text{, }\theta
_{0}\right)  \in%
\mathbb{R}
_{+}\times%
\mathbb{R}
_{+}:f_{\mathcal{P}}\left(  \lambda\text{, }\theta_{0}\right)  <0\right\}  $.
A plot of such a region $\mathcal{P}$ appears in Fig. $3$. We emphasize that,
for values of the parameter $\lambda$ sufficiently large, the power law decay
strategy outperforms the exponential decay strategy in terms of entropic
speed. To have a physical grasp of a typical value of $\lambda$, we recall
that $\lambda=\left(  4\Gamma\right)  /h$ and $\Gamma=\left(  \left\vert
e\right\vert \hslash B_{\bot}\right)  /2mc$. Therefore, assuming to consider a
magnetic field with initial intensity $B_{\bot}$ of the order of $0.2$
$\mathrm{T}$ (a value typical of neodymium magnets), $\lambda\simeq37$
$\left[  \mathrm{MKSA}\right]  $. Furthermore, in Table III we summarize the
behavior of the entropy production rate\textbf{ }$r_{\mathrm{E}}$\textbf{,
}the entropic speed\textbf{ }$v_{\mathrm{E}}$\textbf{, }and the Fisher
information\textbf{ }$\mathcal{F}\left(  \theta\right)  $\textbf{ }for each of
the four quantum mechanical evolutions considered. Finally, for each quantum
evolution, we specify the type of continuous-time quantum search it resembles.

\begin{table}[t]
\centering
\begin{tabular}
[c]{c|c|c|c|c}\hline\hline
Analog Quantum Search & $\mathrm{su}(2$; $\mathbb{C} )$ Hamiltonian Model &
Fisher Information & Speed & Entropy Production Rate\\\hline
Grover-like & constant $B_{\bot}$, original & constant & higher & higher\\
Grover-like & oscillating $B_{\bot}$, generalized & oscillatory & high &
high\\
fixed-point-like & power law decay of $B_{\bot}$, generalized & power law
decay & low & low\\
fixed-point-like & exponential decay of $B_{\bot}$, generalized & exponential
decay & lower & lower\\\hline
\end{tabular}
\caption{Schematic behavioral description of the entropy production rate,
speed, and Fisher information in the selected four $\mathrm{su}(2$;
$\mathbb{C})$ Hamiltonian models. In particular, for each model, we point out
the Grover-like or fixed-point-like property exhibited by its corresponding
analog quantum search algorithm.}%
\end{table}

\section{Conclusions}

In this article, we presented an information geometric characterization of
entropic speeds and entropy production rates that emerge from the geodesic
motion on manifolds of parametrized quantum states. These pure states emerge
as outputs of suitable $\mathrm{su}\left(  2\text{; }%
\mathbb{C}
\right)  $ time-dependent Hamiltonian evolutions employed to specify distinct
types of continuous-time quantum search schemes. The Riemannian metrization on
the manifold is essentially specified by the Fisher information evaluated
along the parametrized squared probability amplitudes obtained from the
analysis of the quantum mechanical temporal evolution of a spin-$1/2$ particle
in an external time-dependent magnetic field that prescribes the
$\mathrm{su}\left(  2\text{; }%
\mathbb{C}
\right)  $ Hamiltonian model. In Fig. $1$, we show the manner in which a
specific Fisher information behavior arises from a given $\mathrm{su}\left(
2\text{; }%
\mathbb{C}
\right)  $ Hamiltonian model characterized by a particular magnetic field
configuration (see Table II). We use a minimum action principle to transfer a
quantum system from an initial state to a final state on the manifold in a
finite temporal interval. Furthermore, we demonstrate that the minimizing
(optimum) path is the shortest (geodesic) path between the two states and in
particular, also minimizes the total entropy production, that is, the
thermodynamic divergence of the path that occurs during the transfer. Then, by
evaluating the entropic speed and the total entropy production along the
optimum transfer paths in the four chosen physical scenarios of interest in
analog quantum search problems, we demonstrate in a clear quantitative manner
that to a faster transfer there necessarily corresponds a higher entropy
production rate (see Fig. $2$ and Table III). Thus, we conclude that lower
(entropic) efficiency values do appear to accompany higher (entropic) speed
values in quantum transfer processes. In particular, quantum mechanical
evolutions that generate probability paths with a Fisher information that
exhibits an exponential decay behavior seem to achieve the highest entropic
efficiency with the cost of also having the lowest entropic speed. By
contrast, probability paths with a constant Fisher information appear to be
the fastest but also the most inefficient from an entropic standpoint. A
graphical summary of these results, including all four $\mathrm{su}\left(
2\text{; }%
\mathbb{C}
\right)  $ Hamiltonian models considered in this paper (see Table II), appears
in Fig. $2$ and Table III.

In conclusion, we view our investigation presented in this paper as a natural
progression of our works presented in Refs. \cite{cafaropre18,alsing19}. It
constitutes a nontrivial preliminary effort toward understanding quantum
search algorithms from a thermodynamical perspective developed within an
information geometric setting. It is our intention to improve upon the
analysis provided in this paper and pursue these fascinating lines of
investigations in forthcoming scientific efforts. Of course, it is our sincere
hope that our work will inspire other scientists to further explore these
research avenues in the near future.

\begin{acknowledgments}
C. C. is grateful to the United States Air Force Research Laboratory (AFRL)
Summer Faculty Fellowship Program for providing support for this work. Any
opinions, findings and conclusions or recommendations expressed in this
manuscript are those of the authors and do not necessarily reflect the views
of AFRL.
\end{acknowledgments}


\begin{thebibliography}{99}                                                                                               %


\bibitem {grover97}L. K. Grover, Phys. Rev. Lett. \textbf{79}, 325 (1997).

\bibitem {nielsenbook}M.\ A. Nielsen and I. L. Chuang, \emph{Quantum
Computation and Information}, Cambridge University Press (2000).

\bibitem {alvarez00}J. J. Alvarez and C. Gomez, arXiv:quant-ph/9910115 (2000).

\bibitem {wadati01}A. Miyake and M. Wadati, Phys. Rev. \textbf{A64}, 042317 (2001).

\bibitem {cafaro2012A}C. Cafaro and S. Mancini, AIP Conf. Proc. \textbf{1443},
374 (2012).

\bibitem {cafaro2012B}C. Cafaro and S. Mancini, Physica \textbf{A391}, 1610 (2012).

\bibitem {cafaro2017}C. Cafaro, Physica \textbf{A470}, 154 (2017).

\bibitem {bennett82}C. H. Bennett, Int. J. Theor. Phys. \textbf{21}, 905 (1982).

\bibitem {parrondo15}J. M. R. Parrondo, J. M. Horowitz, and T. Sagawa, Nat.
Phys. \textbf{11}, 131 (2015).

\bibitem {carlo13}C. Cafaro and P. van Loock, AIP Conf. Proc. \textbf{1553},
275 (2013).

\bibitem {carlo14b}C. Cafaro and P. van Loock, Physica \textbf{A404}, 34 (2014).

\bibitem {beals13}R. Beals,\ S. Brierley, O. Gray, A. W. Harrow,\ S. Kutin, N.
Linden, D. Shepherd, and M. Stather, Proc. R. Soc. \textbf{A469}, 20120686 (2013).

\bibitem {perlner17}R. Perlner and Y.-K. Liu, arXiv:quant-ph/1709.10510 (2017).

\bibitem {cafaropre18}C. Cafaro and P. M.\ Alsing, Phys. Rev. \textbf{E97},
042110 (2018).

\bibitem {byrnes18}T. Byrnes, G. Forster, and L. Tessler, Phys. Rev. Lett.
\textbf{120}, 060501 (2018).

\bibitem {alsing19}C. Cafaro and P. M. Alsing, Int. J. Quantum Information
\textbf{17}, 1950025 (2019).

\bibitem {alsing19b}C. Cafaro and P. M. Alsing, Physica Scripta \textbf{94},
085103 (2019).

\bibitem {cover06}T. M. Cover and J. A. Thomas, \emph{Elements of Information
Theory}, John Wiley \& Sons, Inc. (2006).

\bibitem {frieden98}B. R. Frieden, \emph{Physics from Fisher Information},
Cambridge University Press (1998).

\bibitem {amari00}S.\ Amari and H. Nagaoka, \emph{Methods of Information
Geometry}, Oxford University Press (2000).

\bibitem {brau996}S. L. Braunstein, C. M. Caves, and G. J. Milburn, Annals of
Physics \textbf{247}, 135 (1996).

\bibitem {caves94}S. L. Braunstein and C. M. Caves, Phys. Rev. Lett.
\textbf{72}, 3439 (1994).

\bibitem {crooks12}G. E. Crooks, http://threeplusone.com/sher (2012).

\bibitem {weinhold75}F. Weinhold, J. Chem. Phys. \textbf{63}, 2479 (1975).

\bibitem {ruppeiner79}G. Ruppeiner, Phys. Rev. \textbf{A20}, 1608 (1979).

\bibitem {salamon84}P. Salamon, J. Nulton, and E. Ihrig, J. Chem. Phys.
\textbf{80}, 436 (1984).

\bibitem {salamon83}P. Salamon and R. S. Berry, Phys. Rev.\ Lett. \textbf{51},
1127 (1983).

\bibitem {salamon85}P. Salamon, J. D. Nulton, and R. S. Berry, J. Chem. Phys.
\textbf{82}, 2433 (1985).

\bibitem {crooks07}G. E. Crooks,\emph{ }Phys. Rev. Lett. \textbf{99}, 100602 (2007).

\bibitem {felice90}F. De Felice and J. S. Clarke, \emph{Relativity on curved
manifolds}, Cambridge University Press (1990).

\bibitem {diosi96}L. Diosi, K. Kulacsy, B. Lukacs, and A. Racz, J. Chem. Phys.
\textbf{105}, 11220 (1996).

\bibitem {crooks17}G. M. Rostskoff, G. E. Crooks, and E. Vanden-Eijnden, Phys.
Rev. \textbf{E95}, 012148 (2017).

\bibitem {beretta05}E. P. Gyftopoulos and G. P. Beretta, \emph{Thermodynamics:
Foundations and Applications}, Dover Publications, Inc. (2005).

\bibitem {anandan90}J. Anandan and Y. Aharonov, Phys. Rev. Lett.\textbf{\ 65},
1697 (1990).

\bibitem {sakurai94}J. J. Sakurai, \emph{Modern Quantum Mechanics},
Addison-Wesley Publishing Company, Inc. (1994).

\bibitem {carlopra10}C. Cafaro and S. Mancini, Phys. Rev. \textbf{A82}, 012306 (2010).

\bibitem {carlopra14}C. Cafaro and P. van Loock, Phys. Rev. \textbf{A89},
022316 (2014).

\bibitem {messina14}A. Messina and H. Nakazato, J. Phys. A: Math. Theor.
\textbf{47}, 445302 (2014).

\bibitem {grimaudo18}R. Grimaudo, A. S. M. de Castro, H. Nakazato, and A.
Messina, Ann. Phys. (Berlin) \textbf{2018}, 1800198.
\end{thebibliography}
\end{document}